# Comparison of the Thermospheric Nitric Oxide Emission Observations and the Global Ionosphere-Thermosphere Model (GITM) Simulations: Sensitivity to Solar and Geomagnetic Activities

Cissi Y. Lin[1]

Yue Deng[1]

Karthik Venkataramani[2]

Justin Yonker[2]

Scott M. Bailey[2]

[1]Department of Physics, University of Texas at Arlington, Arlington, Texas, USA

[2]Center for Space Science and Engineering Research, Virginia Tech, Blacksburg, Virginia, USA

Corresponding author: yuedeng@uta.edu

**Abstract**

An accurate estimate of the energy budget (heating and cooling) of the ionosphere and thermosphere, especially during space weather events, has been a challenge. The abundance of Nitric Oxide (NO), a minor species in the thermosphere, is an important component of energy balance here because its production comes from energy sources able to break the strong bond of molecular nitrogen, and infrared emissions from NO play an important role in thermospheric cooling. Recent studies have significantly improved our understanding of NO chemistry and its relationship to energy deposition in the thermospheric photochemical reactions. In this study, the chemical scheme in the Global Ionosphere Thermosphere Model (GITM) is updated to better predict the lower thermospheric NO responses to solar and geomagnetic activity. We investigate the sensitivity of the 5.3-µm NO emission to F10.7 and Ap indices by comparing the global integrated emission from GITM with an empirical proxy derived from the Sounding of the Atmosphere using Broadband Emission Radiometry (SABER) measurements. GITM's total emission agrees well within ±20% of the empirical values. The updated chemistry scheme significantly elevates the level of integrated emission compared to the previous scheme. The inclusion of $N_2(A)$-related production of NO contributes an additional 5–25% to the emission. Localized enhancement of ~70% in column density and a factor of three in column emission are simulated at a moderate geomagnetic level.

Key Points

1. The chemistry scheme in GITM is updated with modern lab results.
2. GITM's NO global emission agrees within 20% of TCI and shows similar relationship to geophysical indicators of energy deposition.
3. Increases of 70% in column density and factor of three in column emission are observable during a simulated storm.

# 1 Introduction

Nitric Oxide (NO) plays several key roles in the lower thermosphere albeit being a minor constituent. Its production is sensitive to those energy sources able to break the strong bond of molecular nitrogen, and its concentrations are indicative of the form of local energy deposition, such as energetic particle precipitation at high latitudes [Baker et al., 2001; Barth et al., 2001] and solar soft X-ray at equatorial latitudes [Barth et al., 1988; 1999; 2003; Barth and Bailey, 2004; Siskind et al., 1990]. Cooling through infrared emission at 5.3 µm [Kockarts, 1980], NO is a crucial part of the thermospheric energy balance. Given that NO has the lowest ionization potential in the chemical reaction chain in the thermosphere, $NO^+$ is the terminal ion and the dominant source of electrons in the E-region of ionosphere. When NO is transported to lower altitudes, such as during sudden stratospheric warming [Bailey et al., 2014; Randall et al., 2009], it is a catalytic destroyer of ozone [Crutzen, 1970; Solomon et al., 1982].

Given its sensitivity to energy deposition, NO densities show the signatures of various spatial and temporal scales, including 11-year solar cycles [Kumar et al., 1999], 27-day solar rotation [Barth et al., 1999], hours-long flares [Rodgers et al., 2010] and enhanced particle participation [Baker et al., 2001]. The effect of solar soft X-ray on thermospheric NO was first proposed by Barth et al. [1988] and Siskind et al. [1990]. A 1D photochemical model suggested that NO density at its peak is a linear function of solar soft X-ray but nonlinear to particle precipitation [Barth, 1992]. A time-dependent thermospheric model [Bailey et al., 2002] later showed that solar soft X-ray is the main driver for the variation of NO density at the equator and particle precipitation is the main driver at high latitudes, as observed by the polar-orbiting Student Nitric Oxide Explorer (SNOE) satellite [Barth et al., 2003].

Since the launch of the Thermosphere-Ionosphere-Mesosphere Energetics and Dynamics (TIMED) satellite in 2001, the onboard Sounding of the Atmosphere using Broadband Emission Radiometry (SABER) instrument has provided valuable measurements for a broad spectrum of aeronomical studies and helped answer a number of questions regarding atmospheric dynamics and chemistry. Based on 14-year SABER measurements, the Thermosphere Climate Index (TCI), parameterized through F10.7, Ap, and Dst indices, was proposed as an empirical proxy for global NO emission [Mlynczak et al., 2015; 2016]. During geomagnetic storms, enhanced emission has been observed to counterbalance the sudden increase of energy deposition in the upper atmosphere [Mlynczak et al., 2003; Knipp et al., 2017; Lei et al., 2012]. However, modeling efforts in general continue to face challenges in reproducing the observed geophysical energy input factors, temperature, and the SABER-observed NO emission in the fundamental vibrational-rotational band at 5.3 µm [Lu et al., 2010]. During storm periods, modeled NO cooling is weaker compared to the SABER emission and the associated neutral density variations deviate up to 60% from those measured along satellite trajectories [Sheng et al., 2017]. The integrated modeled NO cooling power over polar caps appears insufficient to balance the energy from the high-speed streams [Verkhoglyadova et al., 2016]. The quality of representations of the spatial and temporal distribution of the energy drivers and photochemical reactions play parts in these inaccuracies in predicted thermospheric temperatures and densities, in this study, we focus on the photochemical scheme in the model.

In this study, new chemical reactions are added and rates updated to the employed Global Ionosphere Thermosphere Model (GITM) and the sensitivity of NO cooling to the solar and geomagnetic activities are examined before and after the changes. As one of the challenges models face is to provide consistent and reliable skill scores in prediction or reproduction at various activity levels. We intend to utilize the proxy TCI to validate GITM NO cooling in a climatological sense. The goal is to improve the model in regard to both climatological and event studies on thermospheric energy budget. The fundamental NO chemical scheme in the thermosphere is described in Section 2, and the detailed updates are tabulated in Tables 1–3. We note that the revisions studied in this work only serve as an update to the current GITM chemistry scheme [Ridley et al., 2006].

An important update to the NO photochemistry here is the inclusion of the electronically excited state of nitrogen, $N_2(A)$, as an additional source of NO [Guerra et al., 2001; Ono and Oda, 2002; Rousseau et al., 2005; Ionikh et al., 2006; Gatilova et al., 2007; Kutasi et al., 2007; Pintassilgo et al., 2009; Loureiro et al., 2011], implemented as suggested by Yonker [2013]. An in-depth discussion of our current understanding of NO photochemistry in the context of a 1D model can be found in [Yonker, 2013]. In the present work, we perform 3D global simulations under three different chemistry schemes to obtain global integrated emission at 5.3 µm at four solar and geomagnetic levels. We evaluate the sensitivity of model output to these forcing terms against the values calculated from the empirical equation. At the end, we take a glance at the storm response with the new chemistry scheme.

## 2 NO Chemistry in the Thermosphere

The process of NO production and loss has long been a topic of study [Barth, 1992; Bailey et al., 2002, and references therein]. It is produced by the oxidation of atomic nitrogen. The following description is summarized from Yonker [2013].

$$N(\,^2D,\,^4S) + O_2 \rightarrow NO + O \tag{R1}$$

Excited atomic nitrogen, N($^2$D), is the primary source of NO near the NO peak at ~105 km because the oxidation of ground-state atomic nitrogen, N($^4$S), is inefficient at the temperatures there near 300K. At higher altitudes, the importance of N($^4$S) increases as higher temperatures increase the rate of N($^4$S) oxidation much more significantly than that of N($^2$D). Both atomic nitrogen sources of R1, N($^2$D) and N($^4$S), are supplied by photodissociation of N$_2$.

$$N_2 + h\nu \rightarrow N(\,^2D,\,^4S) + N(\,^2D,\,^4S) \tag{R2}$$

NO is mainly destroyed by the cannibalistic reaction (R3) and also charge exchange with O$_2^+$ (R4).

$$NO + N(\,^4S) \rightarrow N_2 + O \tag{R3}$$

$$NO + O_2^+ \rightarrow NO^+ + O_2 \tag{R4}$$

One of the major sources of N($^4$S) is the photodissociation of NO itself by far ultraviolet (FUV) sunlight [Minschwaner and Siskind, 1993].

$$NO + h\nu \rightarrow N(\,^4S) + O \tag{R5}$$

Ionization of NO by Lyman-α at 121.6 nm (R6) is a loss, but subsequent dissociative recombination of NO$^+$ yields N($^2$D) which ultimately leads to the reformation of NO (R7). The produced N($^2$D), has a very short life time and tends to react with O$_2$ as the main source of NO production in the lower thermosphere. Ultimately the key to understanding the production of NO is to understand the sources of atomic nitrogen and particularly N($^2$D).

$$NO + h\nu \rightarrow NO^+ + e^- \tag{R6}$$

$$NO^+ + e^- \rightarrow N(\,^2D) + O \tag{R7}$$

The ionization rate of molecular nitrogen in the lower thermosphere, which consequently contributes to NO production, is dominated by short-wavelength photons with energy greater than 100 eV [Yonker, 2013]. At the lower thermosphere, ionization of $N_2$ is important as the resultant $N_2^+$ can react with atomic oxygen, forming $NO^+$ and $N(^2D)$.

$$N_2^+(X) + O(^3P) \rightarrow NO^+(^1\Sigma^+) + N(^2D) \tag{R8}$$

Both products of R8 eventually lead to NO production (via R7 and R1 respectively). Dissociation of $N_2$ by photons and photoelectrons (R2) is important to NO production. Since the resultant nitrogen atoms contribute to both production (R1) and loss (R3) of NO. In contrast, the ionization of $N_2$ through R8 leads to the production of two $N(^2D)$ atoms (one directly from R8 and the other one from the resulting $NO^+$ through R7), which quickly leads to the production of two NO molecules through R1. Thus, photoelectron ionization turns out to be the more important process for producing NO than photoelectron dissociation of $N_2$ in the lower thermosphere [Yonker, 2013].

Furthermore, the electronically excited state of nitrogen, $N_2(A)$, has been recognized in the recent decades as a significant contributor to thermospheric NO production through R9 owing to its long radiative lifetime in the low-pressure environment in the thermosphere [Guerra et al., 2001; Ono and Oda, 2002; Rousseau et al., 2005; Ionikh et al., 2006; Gatilova et al., 2007; Kutasi et al., 2007; Pintassilgo et al., 2009; Loureiro et al., 2011].

$$N_2(A) + O(^3P) \rightarrow NO + N(^2D) \tag{R9}$$

This reaction is particularly important as it leads to the direct production of a NO molecule (instead of $NO^+$ in R8) as well as a $N(^2D)$ atom, which will also eventually produce an NO molecule. Yonker [2013] provided a parameterization for $N_2(A)$ production rate and demonstrated with a 1D model, NOx1D [Cleary, 1986; Bailey et al., 2002; Yonker, 2013], the importance of $N_2(A)$ (assuming photochemical equilibrium) to the thermospheric chemistry, particularly the production of NO. When including $N_2(A)$, NO density increases ~25–40% at 110 km and ~10% at 150 km for year 1999 depending on the solar EUV spectra (the empirical spectrum from EUVAC and the measurements from SNOE) used by the 1D model.

## 3 Methodology

### 3.1 GITM with a New Photochemical Scheme

GITM is a self-consistent 3D ionosphere-thermosphere model [Deng et al., 2007; Ridley et al., 2006] with the advantages of flexible 3D grid sizes and the ability to solve for non-hydrostatic solutions. It uses altitudinal grids rather than pressure grids as most GCMs do. In this study, global simulation is performed with horizontal resolutions of 5° in latitude and 5° in longitude. The altitude range extends from 100 to ~600 km with vertical resolution of one third of the scale height. The lower boundary conditions of GITM are specified as follows: neutral species densities from the Naval Research Laboratory Mass Spectrometer Incoherent Scatter Radar (NRLMSISE-00) model [Picone et al., 2002], electron density from the International Reference Ionosphere model [Rawer et al., 1978l; Bilitza et al., 2011], and neutral wind from the Horizontal Wind Model [Drob et al., 2015]. The lower boundary conditions of NO density are empirically set to reflect its latitudinal and local-time dependences as shown in previous studies [Bailey et al., 2002]. The EUVAC model [Richards et al., 1994] specifies solar spectrum. The Weimer [1996] high-latitude electric potential pattern and Fuller-Rowell and Evans [1987] particle precipitation pattern specify geomagnetic energy sources.

As the $N_2(A)$-related NO production can potentially be significant [Yonker, 2013], we include $N_2(A)$ into GITM and update the GITM photochemical scheme following the parameterization proposed by Yonker [2013] based on their work on NOx1D. Tables 1–3 list the updates to GITM, including: new reactions, updates to reaction rates, and branching ratios [references tabulated]. Note that the listed reactions are only a subset of the full chemical scheme in GITM [Ridley et al, 2006]. For example, the Duff et al. [2003] rate coefficient for the N($^2$D) reaction in R1 investigated by Sheng et al. [2017] has already been in GITM so the reaction is not particularly listed in these tables. To validate the outcome involving the N$_2$(A) parameterization scheme, we compare the results from GITM and NOx1D in Section 4. A short introduction to NOx1D thus follows.

NOx1D is specifically designed to model thermospheric NOx, a general term for the odd nitrogen family comprised of N and NO, in the vertical direction. The model uses MSIS vertical profiles of major species and ionization and dissociation rates from the two-stream Global airglow (GLOW) photoelectron model [Solomon et al., 1988; Solomon and Abreu, 1989], where photoelectrons travel along the Earth's magnetic field lines, to solve for the steady-state NO photochemistry. The lower boundary is at 70 km, where N($^4$S) is assumed to be in photochemical equilibrium and NO is given a constant mixing ratio of 12 ppm. At the upper boundary of 250 km, N($^4$S) and NO are assumed to be in diffusive equilibrium. The details of the model setting and governing equations can be found in Bailey et al. [2002]. NOx1D considers vertical molecular and eddy diffusion; however, it does not include the effects of winds or waves. Consequently, the temporal-spatial evolution of dynamic changes is not well simulated. Such transport has been included in the GCMs and is significant especially during solar or geomagnetic events when localized impacts extend to wider regions.

Photoelectrons carry residual energy and can initiate sequential reactions, including ionizing and dissociating O, O$_2$, and N$_2$. This effect was not included in previous version of GITM. The ratio of photoelectron impact ionization to direct photoionization, *pe/pi*, is now parameterized through ionization rates of individual species [Solomon and Qian, 2005]. Yonker [2013] adopted their photoelectron ionization rate of $N_2$ to parameterize the level-specific coefficients for $N_2(A)$ production rate. The $N_2(A)$ production channel of NO by energetic particles is also parameterized in a similar manner as that by photoelectrons. Adopting both photoelectron and $N_2(A)$ parameterization schemes also requires a systematic revision of corresponding coefficients, cross sections, and branching ratios as a function of wavelength, which are also incorporated in GITM through this work. Overall, the following changes have been implemented in GITM for the work presented here: 1) accounting for photoelectron impacts to major thermospheric species; 2) including a new NO production channel through $N_2(A)$; and 3) updating the chemistry scheme to include new reactions and values for reaction rates and branching ratios that reflect our current understanding. For each of geophysical conditions examined in this study, GITM simulations are run with three chemistry schemes: (C1) old chemistry, (C2) new chemistry, and (C3) new chemistry with the inclusion of $N_2(A)$ production channel for NO. The difference between two simulation results indicates primarily the variations from the conditions changed.

### 3.2 Empirical Formula from SABER Data

TCI is a convenient multiple linear regression fit of the SABER-observed global thermospheric NO emission using the F10.7, Ap, and Dst as arguments [Mlynczak et al., 2016]. This result allows for a straightforward comparison between modeled and observed NO emission for different levels of solar and geomagnetic activity. The empirical calculation for TCI uses 60-day averaged F10.7, Ap, and Dst as given in Equation 1.

$$TCI = a_0 + a_1 \times F_{10.7} + a_2 \times A_p + a_3 \times Dst \qquad (1)$$

where the TCI is in units of $10^{11}$ W, Ap and Dst in nT, and $F_{10.7}$ in solar flux units (sfu, 1 sfu = 1 $\times$ $10^{-22}$ Wm$^{-2}$Hz$^{-1}$). The coefficients $a_0$, $a_1$, $a_2$, and $a_3$ are -1.024, $1.562 \times 10^{-2}$, $4.637 \times 10^{-2}$, and $-4.013 \times 10^{-3}$ respectively. The 60-day window is chosen to include full local time samplng by the spacecraft. The TCI computed from Equation 1 and the SABER data show a high correlation of 0.985, but differences of ~20% are sometimes observable, in Figure 4 of Mlynczak et al. [2016]. The differences between the observations and fitted values, however, are not further discussed in the work. They may or may not be physical. It is likely that the 60-day averaging only allows to capture features that have similar or greater temporal scale and consequently that the observed variation of NO emission may not be fully expressed linearly with respect to these three indices. Nevertheless, for a simple empirical formula that facilitates the long-term estimation on NO emission over a wide range of solar activity, the differences of 20% during some specific period of time is reasonably acceptable. Given that TCI is a linear combination of 60-day averaged indices, global simulations performed in this study will be compared to it only in a climatological way rather than during episodic events.

### 3.3 Simulation Setup

Four vernal equinox scenarios featuring *60-day averaged* indices are conducted to represent low/high solar and geomagnetic activities. The solar activity level is specified by F10.7 of 70 and 200 for low and high solar activity respectively. TCI is a function of averaged Ap and Dst, whose variability are 3 < Ap < 34 and -44 < Dst < 5 in 2002–2015 [Mlynczak et al., 2015; 2016]. With the similar level of variability in the two indices, Equation 1 shows that TCI is about 10 times more sensitive to Ap than to Dst. Therefore, we focus our geomagnetic aspect of comparison on the Ap levels. In GITM, the geomagnetic activity level is characterized by Hemispheric Power (HP) [Fuller-Rowell and Evans, 1987] and the north-south component of interplanetary magnetic field (IMF), $B_z$, [Weimer, 1996]. In this study, IMF $B_y$ has been fixed to zero to simplify the condition. The conversion from Ap to Kp is conventional but nonlinear [Rostoker, 1972] and that from Kp to HP (in GW) follows the empirical formula [Zhang and Paxton, 2008] as given in Equation 2.

$$HP = 38.66 \times \exp(0.1967 K_p) - 33.99 \text{ for } K_p \leq 5.0 \quad (2a)$$

$$HP = 4.592 \times \exp(0.4731 K_p) + 20.47 \text{ for } K_p > 5.0 \quad (2b)$$

The pairs of HP and $B_z$ are (13.07, -0.01 nT) and (50.90, -7.0 nT) for low and high geomagnetic level respectively as tabulated in Table 4 along with the corresponding values of Ap and Kp. Again, these values are selected to represent 60-day average solar and geomagnetic conditions. The GITM global simulation is run for three days as the pre-condition case and the changes in solar and geomagnetic conditions are made for the consecutive two days. The quasi-steady state is considered as the system response. The extension of interpreting the quasi-steady state regarding the simulation setup is further discussed in the next section.

**4 Results and Discussions**

**4.1 Cross-Comparison**

In this section, GITM NO outputs with the updated chemistry scheme (C3) are compared with an empirical model and SABER measurements. Comparison among different chemistry schemes will be discussed in Sections 4.2 and 4.3. First, the implementation of the $N_2(A)$ parameterization from GITM is compared with the NOx1D results. Second, NO density and NO emission from GITM are compared against the empirical NO model, NOEM, [March et al., 2004] and the SABER measurements [Mlynczak et al., 2005 and references therein]. Eigenanalysis was applied to SNOE measurement to construct NOEM. The empirical orthogonal functions provides a convenient estimation of zonal-mean NO density with given F10.7, Kp, and day of year. SABER provides ~1500 limb-scanned profiles with ~15 orbits per day in its Level 2A data [http://saber.gats-inc.com/]. The corrected NO emission is used for comparison [Mlynczak et al., 2005].

To investigate the effects of background atmospheric conditions on the $N_2(A)$ parameterization, we first compare the $N_2(A)$ abundance parameterized through photoelectron ionization rate of $N_2$ from the NOx1D and GITM 3D models during geomagnetic quiet time with F10.7 = 200 as an example. The NOx1D model requires 7 days to achieve steady-state conditions. The solar and geomagnetic activity levels are kept constant with the vernal equinox of year 2000 condition during the period of simulation. It is assumed that $N_2(A)$ is in photochemical equilibrium in Yonker's [2013] parameterization, in which its production rate is parameterized through photoelectron ionization rate of $N_2$ and its loss frequency is equal to the sum of chemical loss to $O(^3P)$, $O_2$, and radiative loss in the Vegard-Kaplan band. Figure 1a shows the vertical profile of $N_2(A)$ at the equator with varying local hours from NOx1D. The peak density is $1.1 \times 10^9$ m$^{-3}$ at ~150 km at noon. A 3D simulation is performed under the same solar and geomagnetic conditions. The meridional (Figure 1b) and zonal (Figure 1c) variations from GITM show that its peak density, $4.2 \times 10^9$ m$^{-3}$, occurs at ~150 km and local noon, which is in general consistent with the 1D results from NOx1D. The peak value difference is related to the vertical profiles of $O(^3P)$ and $O_2$. The slight differences in Figure 1 are anticipated. Even though the solar and geomagnetic forcing is kept the same, the 1D and 3D simulations do not have identical vertical $O(^3P)$ and $O_2$ profiles as MSIS profiles refreshes at every time step in NOx1D and GITM solves for these species self-consistently. Through the parameterization, the difference results in a factor of 4–5 greater $N_2(A)$ density observed in the 3D cases. The differences observed in the MSIS and GITM profiles require further investigation. All of the simulations presented in this work are performed at the vernal equinox to represent an average geo-condition.

With the C3 chemistry scheme implemented in GITM, Figure 2 shows the comparisons at 60°N at vernal equinox noon: (a) the GITM NO vertical density compared with NOEM, and (b) the GITM NO cooling compared with the SABER 5.3-µm emission. In Figure 2(a), the peak altitude of GITM profile, shown in red, is at ~106 km and that of the NOEM profile, shown in black, is at 110 km. GITM density is mostly greater than NOEM density by up to ~10% throughout this altitudinal range. The two profiles agree well in terms of magnitude and the peak altitude. NOEM is constructed with 3 empirical orthogonal functions based on 2.5 years of SNOE measurements. At 60°N, the NOEM modeled density differs from the averaged SNOE measurements by more than 30% below 110 km and above 120 km [Marsh et al., 2004]. Meanwhile, the zonal-mean of the resulting GITM NO cooling, shown in red, from the same simulation is compared with the zonal-mean SABER 5.3-µm emission, shown in blue, in Figure 2(b). The 191 SABER emission profiles satisfying the latitudinal (55–65 °N) and local time (10:00–14:00) ranges used to derive the zonal mean profile are shown as light blue dashed lines. The GITM cooling profile peaks at ~130 km while the SABER zonal-mean profile peaks ~5 km higher. The peak altitude of SABER emission profiles, however, varies around this by up to 10–20 km. GITM cooling is slightly higher than SABER measurements at 100–120 km and lower at the higher altitudes. Most GCMs, including GITM, consider for radiative cooling only the removal of kinetic energy from the atmosphere by the NO (1–0) transition. The cooling rate is a strong function of NO and O densities and especially temperature [Kockarts, 1980]. The rate coefficient in this study is $2.59 \times 10^{-11}$ cm$^3$ s$^{-1}$ [Caridade et al., 2008]. The observed SABER NO emission, however, also contains contributions from 3–2 and 2–1 transitions [Mlynczak et al., 2005], which may contribute partially the smaller values in the GITM results when compared to the SABER profiles as shown in Figures 2(b). Without density measurements on the date of comparison, the NOEM profile in Figure 2(a) offers a glance of likely NO vertical profile. Below 110 km, GITM has higher density than NOEM, which can result in higher NO cooling (given the same O density and temperature). Above 120 km, though having higher density, GITM has a lower zonal-mean cooling than SABER. This can be explained by slower temperature growth to exospheric temperature (compared to MSIS) we see in several GITM runs. In general, our simulation results show reasonable agreements with both measured NO density and emission, which indicates that the proposed chemistry scheme in GITM can provide a reliable framework for our further implementation to climatological studies on NO. A more systematical data-model comparison is out the scope of this paper and will be the focus of our subsequent papers.

**4.2 Model Sensitivity**

Figure 3 shows the sensitivities of GITM simulated global NO emission to solar/geomagnetic activities with different chemistry schemes and compares those sensitivities with SABER-derived TCI. Figures 3(a) and 3(b) show the sensitivity to Ap index during solar minimum ($F10.7_{60\text{-day}} = 70$) and solar maximum ($F10.7_{60\text{-day}} = 200$) conditions. Figures 3(c) and 3(d) show the sensitivity to F10.7 during geomagnetic low (HP = 13.07 GW) and geomagnetic high (HP = 50.90 GW) conditions. Black lines represent TCI from the empirical formula with the dashed lines marking ±20% from it, green lines represent GITM simulated global emission with (C1) the old chemistry scheme, blue lines represent GITM simulated global emission with (C2) the new chemistry scheme, and red lines represent GITM simulation with (C3) both the new chemistry scheme and $N_2(A)$ parameterization. Clearly significant increase of integrated emission is observed with the change of chemistry rates from the old chemistry scheme (green, C1) to the new chemistry scheme (blue, C2). The comparison between the blue (C2) and red lines (C3) illustrates that the contribution of $N_2(A)$ is positive in all cases. During geomagnetic quiet time, $N_2(A)$ contributes additional total emission of ~25% at solar minimum and ~14% at solar maximum as shown in Figure 3(c). This provides enhancement in NO emission at high latitudes during geomagnetic events and leads to enhancement of total emission by 18% at solar minimum and 6% at solar maximum as shown in Figure 3(d). With the new chemistry schemes (both C2 and C3), the total emission agrees well within 20% of the TCI values in most of the simulated conditions. While the ±20% different threshold that we choose to evaluate may not be ideal, it provides a simple guild line for this first attempt to compare modeled NO emission to the empirical formulation. The additional $N_2(A)$ production channel provides better agreement with the TCI sensitivity in Figure 3(c). The GITM simulations with the C3 chemistry scheme show NO density compared well to the NOEM and NO emission compared well to the SABER data as in Figure 2. The C3 chemistry scheme will be the main scheme we implement for our current and future works.

GITM's sensitivity to geomagnetic activities is lower (slope = $1.44 \times 10^{-2}$) than TCI's ($a_2 = 4.07 \times 10^{-2}$) at solar minimum as shown in Figure 3(a) but higher (slope = $4.38 \times 10^{-2}$) at solar maximum as in Figure 3(b). GITM sensitivity to solar photon activity shows good agreement with TCI's ($a_1 = 1.56 \times 10^{-2}$) being only negligibly lower (slope = $1.52 \times 10^{-2}$) during geomagnetic quiet time as in Figure 3(c) and slightly higher (slope = $2.38 \times 10^{-2}$) during storm time as in Figure 3(d). Equation 1 shows that TCI has a constant sensitivity to F10.7 and Ap. GITM's sensitivity to solar/geomagnetic activities grows with geomagnetic/solar activities, which indicates some non-linear effects in the simulation. As solar activity rises, the background atmosphere becomes warmer, leading to higher NO density and emission, which contributes to the higher sensitivity of NO emission to geomagnetic activity as shown in Figures 3(a) and 3(b). The coupled ionosphere/thermosphere is a non-linear system by nature. As the deposited energy is intensified during geomagnetic events, thermospheric density, temperature and compositions undergo the integrated effects of the physical and chemical processes, which may not be captured by a linear function of indices.

The interplanetary magnetic field, particular $B_z$, is found to result in the most dynamic variability in GITM TCI among the solar and geomagnetic forcing used in GITM. Several test cases during solar maximum and minimum reveals that the averaged GITM TCI of 60-day simulations with time-varying solar and geomagnetic forcing falls within ±20% of proxy TCI. Depending on how $B_z$ varies over the 60-day period, a departure of up to ~$0.3 \times 10^{11}$ GW from the steady-state GITM TCI obtained using 60-average indices has been observed. This signifies the role $B_z$ plays in the nonlinear IT system in this regard. The $0.3 \times 10^{11}$-GW difference is about 10–20% of TCI on average.

Other factors may contribute to the differences in sensitivity observed between the calculated TCI and GITM model results. As stated earlier, the calculation of NO cooling in GITM considers only the 1–0 transition while the volume emission in the SABER data contains also 3–2 and 2–1 transitions [Mlynczak et al., 2005]. As TCI is a proxy based on multiple linear regression of the SABER measurements, the covariance coefficients that are expected to exist among the indices, F10.7, Ap, and Dst, are not available in its formation for us to further adjust the input forcing levels more coherently. Besides instrumental and retrieval uncertainties residing in data sets, it is observable in Figure 4 of Mlynczak et al. [2016] that the fitting may sometimes deviate from measurements by ~20% for a certain period of time (up to ~6 months) though the overall correlation coefficient is 0.985 over the 14-year observations. On the other hand, direct (such as temperature) and indirect (such as transport) influence from different components in GITM may contribute to the different sensitivities observed in the modeled emission when compared to TCI. The sensitivities to solar and geomagnetic activity levels reflect the combined effect residing in GITM by these components.

When F10.7 increases from 70 to 200, emission increases by a factor of 5.1, as shown in Figure 3(d). However, the total emission increases by 1.8 when Ap increases from 4 to 27 at solar maximum as shown in Figure 3(b). It suggests that the solar irradiation dominates the variation of total emission, which can be explained in both total energy and energy distribution. As pointed out in [Knipp et al., 2004], on average, the solar irradiation and geomagnetic energy inputs contribute 80% and 20% of the total energy budget of upper atmosphere. The partition between solar and geomagnetic contributions to TCI appears to have solar-cycle dependency – solar component varying almost in-phase with solar cycle between 50–90% and geomagnetic component varying almost out-of-phase with solar cycle between 10–50% [Mlynczak et al., 2015]. Meanwhile, the two energy sources have different and distinguishing spatial distributions in latitude and local time. A multiple regression performed on the SNOE measurements shows that the solar X-ray radiation contributes to NO production between 0–60°N and the auroral particles contribute at 25-65°N [Barth et al., 2003].

### 4.3 Density and Emission

Figure 3 shows clear enhancement in total emission at each update of the chemistry scheme: an increase of up to a factor of three from C1 to C2 and 5–25% with the inclusion of $N_2(A)$ production from C2 to C3. In general, changes made in the new chemistry scheme lead to higher production rate of NO and lower loss rate at all altitudes (particularly between 100–200 km). The N($^2$D) production channel in R1 contributes the most to the production increase. Loss through both R3 and R4 weakens in the new chemistry scheme. Figure 4 shows the distribution of modeled NO density at 105 km using (a) old and (b) new chemistry schemes and the (c) percent difference resulting from such chemistry scheme update and (d) from N$_2$(A) with F10.7 = 200, HP = 50.90 GW, Bz = -7 nT. NO density increases of up to 30% are clearly observed throughout the globe with the chemistry updates and up to 10% when including the N$_2$(A) production channel. The dayside and auroral oval, where N$_2$(A) contribution is parameterized through photoelectrons and energetic particles, show slightly greater percentage (>5%) increases. The increase of NO density as shown in Figure 4 consequently results in the increase of total emission observed in Figure 3.

To demonstrate the observable variation in NO in response to energy inputs to the IT system with the new chemistry scheme and N$_2$(A) integrated into GITM, we subtract the low geomagnetic activity level from the high activity level and consider the residual as storm-time response of the thermosphere. Compared to quiet time level, the change of NO column density and emission two hours after HP rises from 13.1 to 50.90 GW and Bz from -0.01 to 7.0 nT is shown in Figure 5. Figure 5a shows that the increase in column density reaches +70% at this point and some degree of depletion is present adjacent to the area of enhancement. The magnitude of enhancement observed in column density agrees well with the cases observed by TIMED/GUVI during geomagnetic storms in 2012 [Zhang et al., 2014]. When the geomagnetic activities intensify, the energy deposition of particle precipitation creates sudden localized chain effects, including composition changes, thermal, and Joule heating. Changes in composition, especially the spatial distribution of NO and O, determine the distribution of NO and O cooling. The disposition of the energy terms also determines temperature, which is the key component not only to chemical reactions involved but also to dynamics in the IT system.

Owing to the significant amount of increase in NO abundance, a strong enhancement in the column emission at the polar and subauroral regions is observed as expected. Figure 5b shows that the maximal enhancement reaches three times of the quiet-time level. The averaged NO 5.3 µm emission rate from TIMED/SABER shows similar magnitude of variability, 1–3 times of pre-storm value, two hours after storm onset regardless in dayside or nightside measurements during two geomagnetic storms in October 2003 [Lei et al., 2012]. As stated previously, in the model calculation NO cooling is a function of NO and O densities and especially temperature [Kockarts, 1980]. The observed spatial distribution in Figure 5 can be explained as modulation of the density distribution – for example, weak emission depletion collocating with weak density depletion in smaller areas – of those two species and temperature. During storm time, the distribution of O changes distinguishingly and is responsible for the observed variation in NO emission.

A separate scenario is performed to study thermospheric response to a stronger geomagnetic storm. The solar and geomagnetic inputs are: F10.7 = 200, HP = 99.0 GW, and Bz = -20 nT. Figure 6 shows the zonal mean NO emission calculated from the 3D GITM simulation two hours after the storm onset. We clarify that the zonal mean is performed on all longitudinal grids (at all local time) simultaneously instead of along particular satellite orbital crossings (for example, TIMED/SABER orbits at noon-midnight for the duration of a storm). The maximum emission of ~$3.5 \times 10^{-8}$ W/m$^3$ occurs between 125 and 150 km. This is consistent with the volume emission rate reported by TIMED/SABER during April 2002 [Mlynczak et al., 2003], which indicates that GITM represents the NO emission response to the geomagnetic activity reasonably well in terms of magnitude and spatial distribution.

**5 Summary and Conclusions**

To establish a reliable computation for global NO emission, we improved the chemistry scheme and incorporated $N_2(A)$ and photoelectron parameterization to GITM, which increases the resulting NO density and 5.3 μm emission. The sensitivities of GITM total NO emission to solar/geomagnetic activities at fixed geomagnetic/solar levels are examined with three chemistry schemes. The updated chemistry scheme enhances the global integrated emission by a factor of up to three and the contribution of $N_2(A)$ adds additional 5–25% of emission in the cases studied. The enhancement in NO emission at high altitudes during geomagnetic events is up to 18% of global integrated emission. With the new chemistry scheme and $N_2(A)$ parameterization, the total emission agrees well within 20% of the empirical TCI values in most cases. NO density at 105 km shows an increase of 30% when the old chemistry scheme is replaced with the new chemistry scheme. The inclusion of $N_2(A)$ production provides additional 5–10% increase in density at the subsolar zone and auroral oval. Such increase by $N_2(A)$ in density is consistent with previous studies.

Impacts of a moderate storm at high solar activity level are also examined preliminarily. The additional energy deposited in the lower thermosphere results in ~70% increase in NO column density and three times greater column emission at the polar region. The magnitudes of these increases in NO column density and column emission and zonal mean emission (up to $3.5 \times 10^{-8}$ W/m$^3$) agree well with TIMED/SABER and GUVI measurements.

**Acknowledgement**

This research at the University of Texas at Arlington was supported by NSF through grant ATM0955629, NASA through grants NNX13AD64G and NNX14AD46G, and AFOSR through award FA9550-16-1-0059 and MURI FA9559-16-1-0364. The authors acknowledge the Texas Advanced Computing Center (TACC) at the University of Texas at Austin for providing Lonestar5 and Maverick resources that have contributed to the research results reported within this paper. URL: http://www.tacc.utexas.edu. The GITM model outputs are available at htttps://github.com/cissilin/Lin2018JGR_GITMNO.

Siskind, D. E., C. A. Barth, and D. D. Cleary (1990), The possible effect of solar soft X rays on thermospheric nitric oxide, J. Geophys. Res., 95(A4), 4311–4317, doi:10.1029/JA095iA04p04311

Solomon, S. C., and V. J. Abreu (1989), The 630 nm dayglow, J. Geophys. Res., 94(A6), 6817–6824, doi:10.1029/JA094iA06p06817

Solomon, S. C., P. B. Hays, and V. J. Abreu (1988), The auroral 6300 Å emission: Observations and modeling, J. Geophys. Res., 93(A9), 9867–9882, doi:10.1029/JA093iA09p09867

Solomon, S., P. J. Crutzen, and R. G. Roble (1982), Photochemical coupling between the thermosphere and the lower atmosphere: 1. Odd nitrogen from 50 to 120 km, J. Geophys. Res., 87(C9), 7206–7220, doi:10.1029/JC087iC09p07206

Solomon, S. C., and L. Qian (2005), Solar extreme-ultraviolet irradiance for general circulation models, J. Geophys. Res., 110, A10306, doi:10.1029/2005JA011160

Verkhoglyadova, O., X. Meng, A. J. Mannucci, B. T. Tsurutani, L. A. Hunt, M. G. Mlynczak, R. Hajra, and B. A. Emery5 (2016), Estimation of energy budget of ionosphere-thermosphere system during two CIR-HSS events: observations and modeling, J. Space Weather Space Clim., 6(A20), doi:10.1051/swsc/2016013

Weimer, D. R. (1996), A flexible, IMF dependent model of high-latitude electric potentials having "Space Weather" applications, Geophys. Res. Lett., 23(18), 2549–2552, doi:10.1029/96GL02255

Yonker, J. D. (2013), Contribution of the first electronically excited state of molecular nitrogen to thermospheric nitric oxide, Ph.D. dissertation

Zhang, Y. and L. J. Paxton (2008), An empirical Kp-dependent global auroral model based on TIMED/GUVI FUV data, J. Atmos. & Solar-Terres. Phys., 70(8–9), 1231–1242, doi:10.1016/j.jastp.2008.03.008

Zhang, Y., L. J. Paxton, D. Morrisona, D. Marshb, and H. Kila (2014), Storm-time behaviors of O/N2 and NO variations, J. Atmos. & Solar-Terres. Phys., 114, 42–49, doi:10.1016/j.jastp.2014.04.003

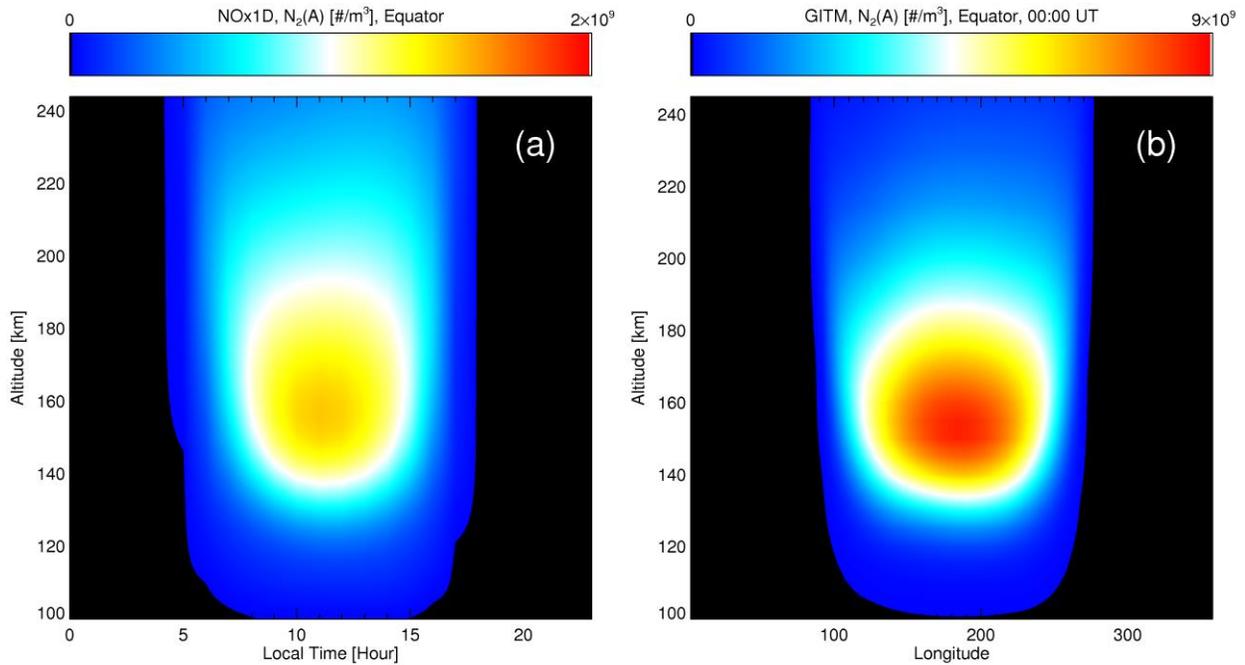

Figure 1. $N_2(A)$ abundance (in the unit of number per cubic meter) parameterized through photoelectron ionization rate of $N_2$ at F10.7 = 200 during geomagnetic quiet time (Ap = 5): (a) 1-D model; GITM (b) zonal and (c) meridional distributions.

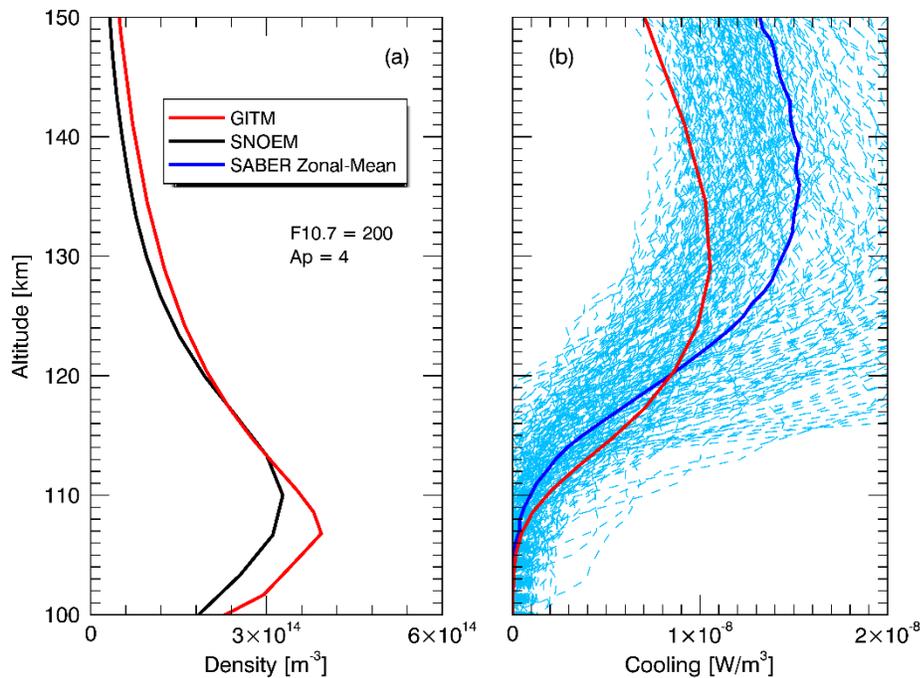

Figure 2. (a) GITM NO density compared with NOEM and (b) NO cooling compared with SABER 5.3-µm emission at the vernal equinox with F10.7 = 200 and Ap = 4. All the SABER emission profiles falling in the latitudinal (55–65 °N) and local time (10:00–14:00) ranges are shown with light blue dashed lines.

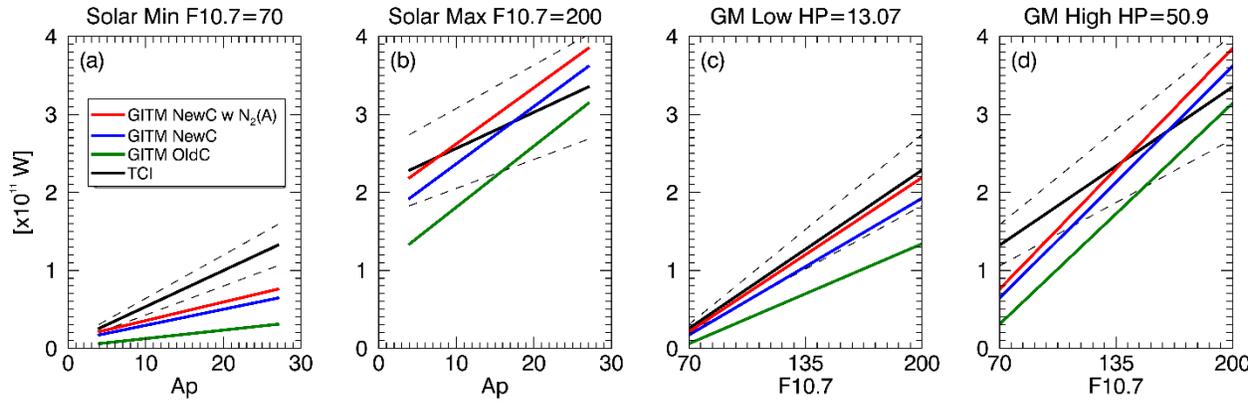

Figure 3. Global emission from three GITM photochemical schemes compared with TCI. GITM's sensitivity to solar/geomagnetic activity level appears to be a function of geomagnetic/solar activity level as the background atmospheric structures differ: sensitivity to Ap at (a) solar minimum (F10.7$_{60\text{-day}}$ = 70) and (b) solar maximum (F10.7$_{60\text{-day}}$ = 200) conditions; sensitivity to F10.7 at (c) geomagnetic low (HP = 13.07 GW) and (d) geomagnetic high (HP = 50.9 GW) conditions. The black lines represent TCI from the empirical equation and the dashed lines mark ±20% from TCI. GITM global emission is shown for (C1) old chemistry scheme in green, (C2) new chemistry scheme in blue, and (C3) new chemistry scheme and N$_2$(A) parameterization in red .

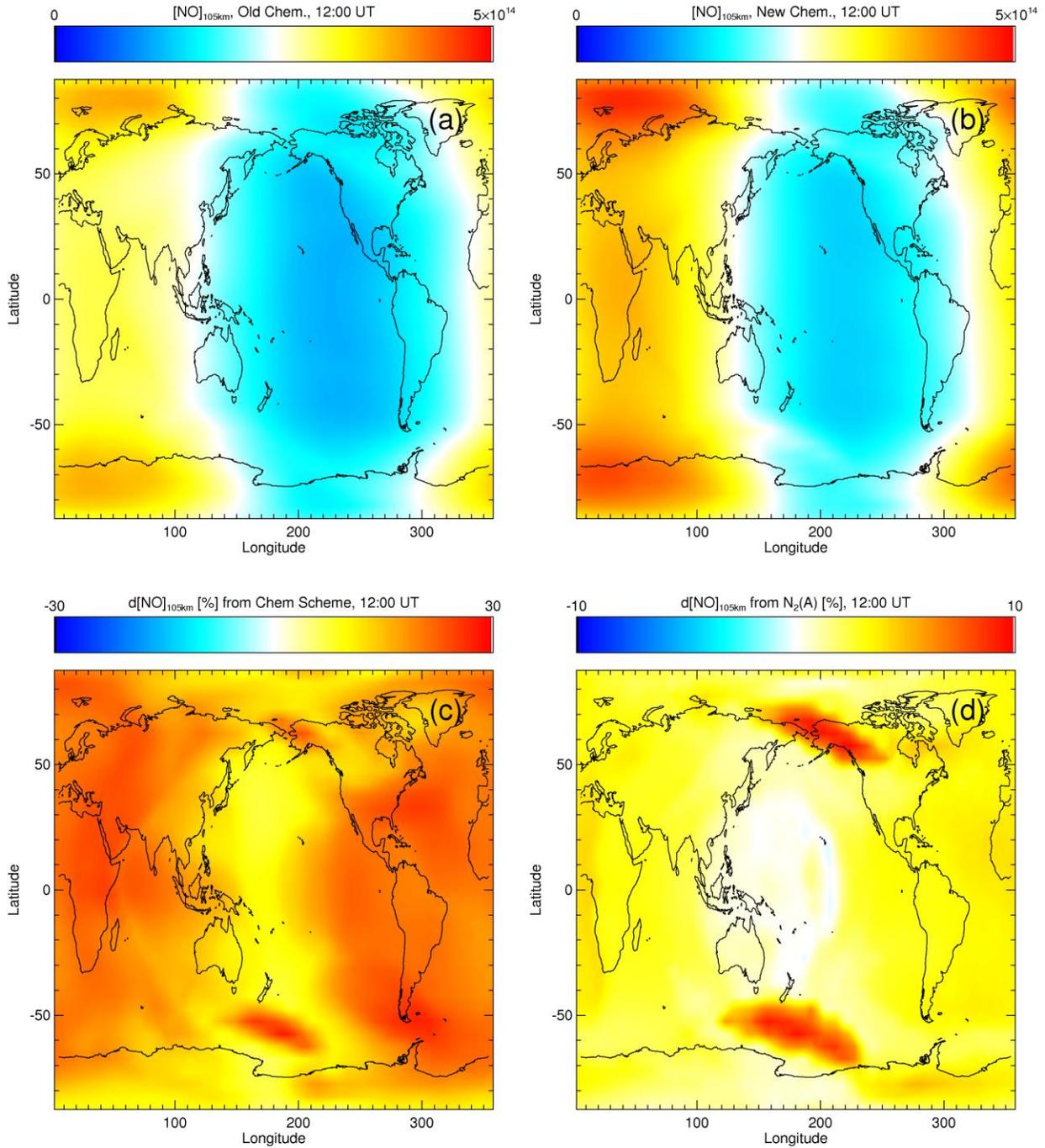

Figure 4. NO volume density at 105 km with the (a) old and (b) new chemistry schemes (without $N_2(A)$) and the percentage differences owing to the update of (c) chemistry scheme (S2 – S1). Similarly, panel (d) shows global increases of NO density result from the inclusion of $N_2(A)$ (S3 – S2). The geophysical conditions are F10.7 = 200, HP = 50.9 GW, and Bz = -7.0 nT.

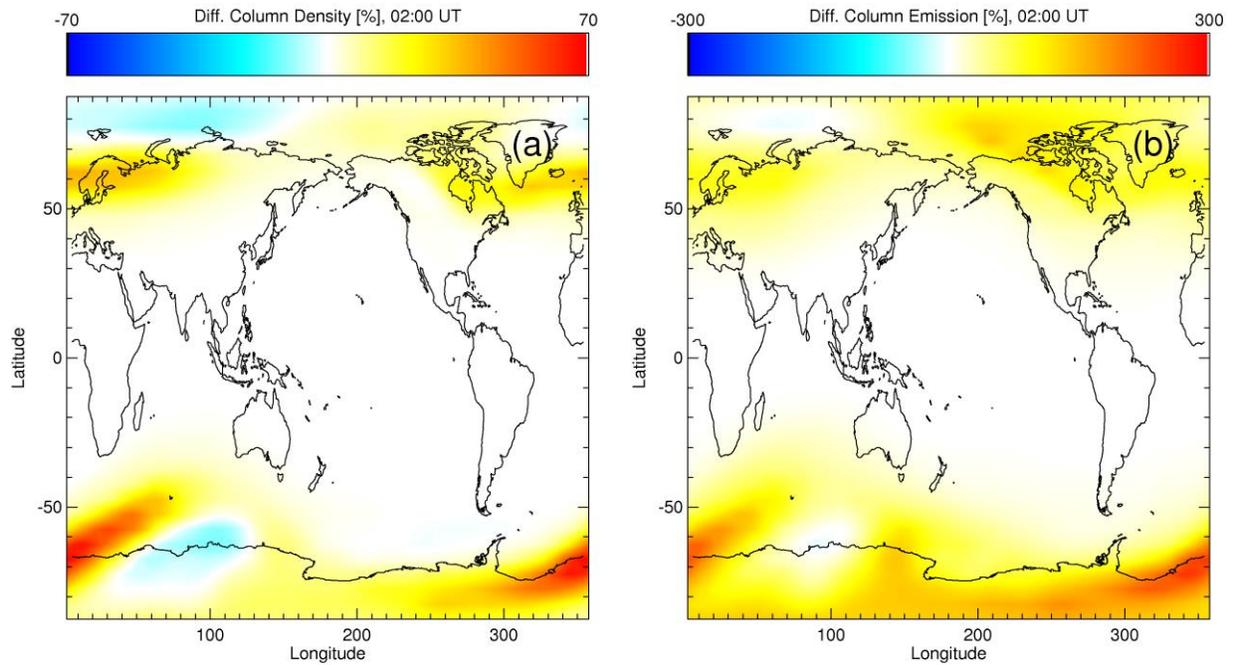

Figure 5. Enhancement in NO column (a) density and (b) emission two hours after a geomagnetic storm onset with both the inclusion of $N_2(A)$ and updated chemistry, with F10.7 = 200, HP = 50.90 GW, and Bz = -7.0 nT.

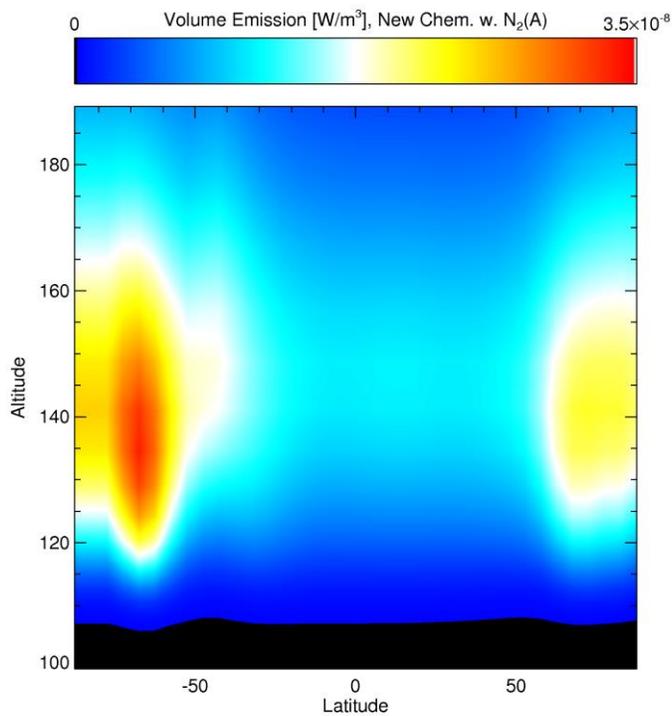

Figure 6. The zonal mean volume emission rate two hours after a geomagnetic storm onset (HP = 99.00 GW, Bz = -20.0 nT).

Table 1: New reactions updated to the GITM chemistry scheme [Ridley et al., 2006]

| Reaction | Reaction Rate | Reference |
|---|---|---|
| $N_2(A) + O(^3P) \longrightarrow NO + N(^2D)$ | $1.0 \times 10^{-19} \cdot \sqrt{\frac{T_e}{298}}$ | Yonker 2013 |
| $N(^2P) + O(^3P) \xrightarrow{0.50} NO^+ + e$ | | |
| $N(^2P) + O(^3P) \xrightarrow{0.47} N(^2D) + O$ | $2.7 \times 10^{-17}$ | Herron, 1999 |
| $N(^2P) + O(^3P) \xrightarrow{0.03} N(^4S) + O$ | | |
| $N(^2D) + N_2 \longrightarrow N(^4S) + N_2$ | $1.74 \times 10^{-20}$ | |
| $N(^2P) + e \longrightarrow N(^2D) + e$ | $9.5 \times 10^{-15}$ | |
| $N(^2P) + e \longrightarrow N(^4S) + e$ | $1.6 \times 10^{-18} \cdot (\frac{T_e}{300})^{0.85}$ | |
| $O_2^+ + N(^2D) \longrightarrow NO^+ + O$ | $1.8 \times 10^{-16}$ | Richards et al, 2011 |

Table 2: Updates of branching ratios and reaction rates to the GITM chemistry scheme [Ridley et al., 2006]

| Reaction | Updated Value | Previous Value | Reference |
|---|---|---|---|
| $N_2^+ + e \xrightarrow{0.52} N(^2D) + N(^2D)$ | | | |
| $N_2^+ + e \xrightarrow{0.37} N(^4S) + N(^2D)$ | same | $2.2 \times 10^{-13} \cdot (\frac{T_e}{300})^{-0.39}$ | Peterson et al 1998, Sheehan & StMaurice 2004 |
| $N_2^+ + e \xrightarrow{0.11} N(^4S) + N(^2P)$ | | | |
| $N_2^+ + O(^3P) \xrightarrow{0.90} N(2D) + NO^+$ | | | |
| $N_2^+ + O(^3P) \xrightarrow{0.05} N(4S) + NO^+$ | same | $1.4 \times 10^{-16} \cdot (\frac{T}{300})^{-0.44}$ | Fox & Sung 2001, Scott et al. 1999, Yonker 2013 |
| $N_2^+ + O(^3P) \xrightarrow{0.05} O^+(^4S) + N_2 + 1.96 eV$ | | | |
| $N^+ + O_2 \xrightarrow{0.50} O_2^+ + N(^2D) + 0.1 eV$ | | | Midey et al. 2006 |
| $N^+ + O_2 \xrightarrow{0.42} NO^+ + O(^3P) + 6.67 eV$ | $5.5 \times 10^{-16}$ | $2.0 \times 10^{-16}$ | Midey et al. 2006 |
| $N^+ + O_2 \xrightarrow{0.08} O^+(^4S) + NO + 2.31 eV$ | | | |
| $N^+ + NO \xrightarrow{0.91} NO^+ + N(^4S) + 3.4 eV$ | | | |
| $N^+ + NO \xrightarrow{0.07} N_2^+ + O + 2.2 eV$ | $6.5 \times 10^{-15} \cdot T_i^{-0.44}$ | $4.72 \times 10^{-16} \cdot (\frac{T_i}{300})^{-0.24}$ | Midey 2004 |
| $N^+ + NO \xrightarrow{0.02} N_2 + O(^4S)^+$ | | | |
| $NO^+ + e \xrightarrow{0.95} N(2D) + O(^3P) + 0.38 eV$ | $3.5 \times 10^{-13} \cdot (\frac{T_e}{300})^{-0.69}$ | $4.2 \times 10^{-13} \cdot (\frac{T_e}{300})^{-0.69}$ | |
| $NO^+ + e \xrightarrow{0.05} N(4S) + O(^3P) + 2.75 eV$ | | | |

Table 3: Updates of reaction rates to the GITM chemistry scheme [Ridley et al., 2006]

| Reaction | Updated Value | Previous Value | Reference |
|---|---|---|---|
| $N(^4S) + O_2 \longrightarrow NO + O + 1.385\,eV$ | $1.5 \times 10^{-17} \cdot exp(-\frac{3600}{T_n})$ | $4.4 \times 10^{-18} \cdot exp(-\frac{3220}{T_n})$ | Sander 2015 |
| $N(^2D) + NO \longrightarrow N_2 + O + 5.63\,eV$ | $6.7 \times 10^{-17}$ | $7.0 \times 10^{-17}$ | Herron 1999 |
| $N(^4S) + NO \longrightarrow N_2 + O + 3.25\,eV$ | $2.1 \times 10^{-17} \cdot exp(\frac{100}{T_n})$ | $1.5 \times 10^{-18} \cdot \sqrt{T_n}$ | Sander 2015 |
| $N(^2D) + O \longrightarrow N(^4S) + O(^3P) + 2.38\,eV$ | $1.65 \times 10^{-18} \cdot exp(-\frac{260}{T_n})$ | $2.0 \times 10^{-18}$ | Yonker 2013 |
| $N_2^+ + O_2 \longrightarrow O_2^+ + N_2 + 3.53\,eV$ | $5.0 \times 10^{-17} \cdot (\frac{T}{300})^{-1.16}$ | $5.1 \times 10^{-17} \cdot (\frac{T}{300})^{-0.8}$ | Fox & Sung 2001 |
| $N_2^+ + NO \longrightarrow NO^+ + N_2 + 6.33\,eV$ | $7.5 \times 10^{-15} \cdot T^{-0.52}$ | $3.3 \times 10^{-16}$ | Midey et al. 2004 |
| $O_2^+ + NO \longrightarrow NO^+ + O_2 + 2.813\,eV$ | $4.5 \times 10^{-16}$ | $4.4 \times 10^{-16}$ | Midey & Viggiano 1999 |
| $O_2^+ + N(^2D) \longrightarrow N^+ + O_2 + 0.0\,eV$ | $8.65 \times 10^{-17}$ | $2.5 \times 10^{-16}$ | Richards 2011 |
| $O_2^+ + N(^4S) \longrightarrow NO^+ + O + 4.21\,eV$ | $1.0 \times 10^{-16}$ | $1.8 \times 10^{-16}$ | Scott et al. 1998 |
| $O_2^+ + e \longrightarrow 2\,O$ | $1.95 \times 10^{-13} \cdot (\frac{T_e}{300})^{-0.7}$ | $2.4 \times 10^{-13} \cdot (\frac{T_e}{300})^{-0.7}$ | Sheehan & St-Maurice 2004 |
| $O^+(^2D) + N \longrightarrow N^+ + O + 1.0\,eV$ | $1.5 \times 10^{-16}$ | $7.5 \times 10^{-17}$ | |
| $O^+(^2P) + e \longrightarrow O^+(^2D) + e + 1.69\,eV$ | $1.84 \times 10^{-13} \cdot (\frac{T_e}{300})^{-0.5} + 3.03 \times 10^{-14} \cdot (\frac{T_e}{300})^{-0.5}$ | $1.5 \times 10^{-13} \cdot (\frac{T_e}{300})^{-0.5}$ | Richards 2011 |
| $N^+ + O(^3P) \longrightarrow O^+ + N(^4S) + 0.93\,eV$ | $4.5 \times 10^{-18}$ | $2.2 \times 10^{-18}$ | Anicich 2003 |

Table 4: Model settings for climatological comparison during low and high geomagnetic activities and the corresponding 60-day averaged geomagnetic indices.

| Geomagnetic Activity Level | Low | High |
|---|---|---|
| Ap | 4 | 27 |
| Kp | 1 | 4 |
| HP [GW] | 13.07 | 50.90 |
| Bz [nT] | -0.01 | -7.00 |